# yNet: a multi-input convolutional network for ultra-fast simulation of field evolvement

Zhuo Wang[1], Xiao Wang[2], Wenhua Yang[3], Yaohong Xiao[1], Yucheng Liu[3], Lei Chen[1,4,*]

[1] *Department of Mechanical Engineering, University of Michigan-Dearborn, Dearborn, MI 48128, USA*
[2] *Department of Mechanical Engineering, Jiangsu Normal University,*
[3] *Department of Mechanical Engineering, Mississippi State University, Starkville, MS 39762, USA*
[4] *Michigan Institute for Data Science, University of Michigan, Ann Arbor, MI 48109, USA*

**Abstract**

The capability of multi-input field-to-field regression, i.e. mapping the initial field and applied conditions to the evolved field, is appealing, enabling ultra-fast physics-free simulation of various field evolvements across many disciplines. We hereby propose a y-shaped multi-input deep convolutional network, yNet, which can effectively account for combined effects of multiple mixed inputs on output field. The proposed yNet is applied to the simulation of porosity evolution in selective lasering sintering (SLS). Upon testing, yNet can simulate nearly identical porosity evolution and development to the physics-based model, with a 99.13% morphological similarity for various SLS conditions. We then effortlessly boost the porosity simulation capability to the realistic, full-component level. yNet is generally applicable to simulating various structural/morphological evolutions and other condition-concerned, continuous field evolvements even with spatially and/or temporally non-uniform evolving kinetics. Once trained, the light-weight yNet can be distributed easily and ran with limited computational resource while reducing computation time to a split second. It thus may have a transformative impact by democratizing the capability of ultra-fast and extreme-scale simulation of various field evolvements.

## 1. Introduction

As a cornerstone of deep learning, deep convolutional network (ConvNet) features utilization of multilayer stack of convolutional layers for automatic, hierarchical representation learning. The four key components of ConvNets, i.e. local connections, shared weights, pooling and the use of many layers, bring it great advantages in processing data in the form of multiple arrays, such as signals and sequences (1D), images or audio spectrograms (2D) and video or volumetric images (3D) [1]. Since its first application in digit recognition nearly 30 years ago [2], it has been especially serving as the backbone for performing various computer vision (CV) or image-centered tasks, such as image labeling/classification [3], object localization [4], semantic segmentation (pixel-wise labeling) [5, 6], etc. Upon realization of its distinct advantage in image processing, it has been recently witnessed a surging application of ConvNets in engineering field where image-involved problems are ubiquitous. Some of those engineering applications include image classification with respect to engineering-related dataset [7, 8], structure-property relationship modeling [9, 10], microstructure characterization and reconstruction (MCR) [11]. They all feature the leverage of ConvNets to explicitly read and process images or microstructures in image form, thereby free from hand-craft featurization and with minimal human intervention.

Besides those applications, another prominent yet technically challenging one might be ConvNet based simulation of field evolvement and development. It can be regarded as physics-free simulation, since ConvNets treat the field evolvement plainly as a (field-to-field) regression problem, irrespective of underlying physics. Some few related researches [12, 13] just adopt the original CV task oriented ConvNets. However, image-to-image/field-to-field regression tasks are rather different in the context of CV and engineering. First, the input of CV tasks is solely image while the field-to-field regression in engineering contains multiple mixed inputs. In CV tasks or specifically semantic segmentation, the output image or segmentation result is uniquely determined by the input image to be segmented. In the context of engineering, the developed field is nonetheless not only decided by the initial field, but also affected by the applied conditions. Due to the technical limit, existing researches simplified the regression task by fixing the condition, thus failing to incorporate condition-related parameters. This is apparently contrary to the common fact of great variability in applied conditions and seriously stifles the wide usage of the trained ConvNets. Second, in semantic segmentation, the input image would essentially share an underlying structure with output image, in contrast to field evolvement/change in engineering problems. So far, developing such a multi-input field-to-field regressor, which can capture the combined effects of multiple inputs of same significance yet with striking format and weight difference (i.e. high-dimensional image versus point values), remains a daunting task.

---

*[leichn@umich.edu](mailto:leichn@umich.edu) (L. Chen)

To address this critical gap, we proposed a multi-input deep convolutional neural network, which is conceptually simple yet powerful for handling multi-input field-to-field regression tasks in engineering. We name it "yNet" based on its signature y-shape (see Fig. 1a). This name conveys its most salient feature, i.e. the merge of an additional flow of input signal from applied conditions, in comparison to the pure image-to-image regression neural network (thus I-shaped) in aforementioned CV and engineering applications. We emphasize that, by multi-input, it means the multiplicity of not only the quantity (initial field and various condition-related parameters) but also the type (high-dimensional image and point values). Also noted is that, the term "field-to-field regression" is preferably used due to the engineering problem targeted nature of yNet; because in engineering what underlies an image is usually a practically or physically meaningful field, such as temperature, stress/strain and velocity fields. It should be stressed that various structures in engineering represent a typical family of field. For instance, in the phase-field model [14] a wide range of structures, such as polycrystals [15], dendrites [16] and precipitate [17], are described using so-called phase field variable, while common solid structures of any geometry may be just represented using a density field, as broadly seen in topology optimization [18-20].

Besides the concept of field, we further clarify applied conditions. Here the umbrella term "condition" refers to not only external conditions (e.g. ambient temperature and applied loading level) but also internal conditions (e.g. materials and physical properties concerned with evolving kinetics), as well as any implicit factor (e.g. laser parameters in this paper) and even the time period of evolvement. As long as one has an association with the field development, either explicitly or implicitly, it can be seen as "condition", thus falling within the interest of yNet.

In the remaining sections, we choose simulation of porosity development in selective laser sintering (SLS) as a case study to instantiate yNet. SLS [21] is one of the most popular additive manufacturing (AM) techniques, widely adopted for fabricating metal, ceramic and polymer components. The specific reason for selecting porosity prediction problem in SLS process is three-fold. First, unlike other AM processes such as selective laser melting (SLM) and laser metal deposition (LMD), sintering-based powder binding is a mild process with insignificant melting&solidification phenomenon, which makes porous structure the priority feature of SLS-fabricated component [22]. A fast yet reliable predictive capability of porosity formation in SLS is of huge practical significance. However, the existing physics-based approaches are notorious for their computational complexity, due to the miscellaneous underlying physics involved. Accurate simulation of porosity formation in SLS usually requires proper consideration of rigid-body translation and rotation of powder particles, grain growth through boundary migration, and a plethora of diffusion mechanisms [23]. This fact makes a substitute of the cumbersome physics-based model a desperate need. Second, the porosity formation essentially results from the evolvement (e.g. rounding, shrinkage and even complete closure) of inter-powder gaps in as-deposited powder bed [24], and the porosity development does depend on a variety of SLS parameters, such as laser power, laser scanning speed and preheating temperature. It thus falls perfectly within the task of interest of yNet, i.e. condition-controlled field evolvement. Third, unlike most structural evolutions, the porosity development is also affected by the distance to the heat source applied at the powder bed surface, thereby making evolving kinetics spatially non-uniform and complex. It opens up good opportunity to examine the robustness of yNet in a tough application scenario.

Next, we will introduce the physical SLS model and elaborate the architecture design of yNet, followed by training and testing of yNet. The trained yNet is then applied to addressing the as-of-yet challenging task of large-scale SLS porosity simulation. Finally, we analyzed contributing factors to various merits of yNet, and discussed its limit as well as the enormous potential in future applications.

## 2. Physical SLS model

Physical SLS simulation provides training and testing data. In this study, we basically extend the primitive phase-field-based sintering model [23] for applicability in SLS, by incorporating heat transfer model and a Gaussian heat source model describing the effective heat input from moving laser beam [25]. The operating range of laser power is [25, 40] W and scanning speed [0.5, 2.5] m s$^{-1}$. Note that they will be normalized in use by yNet. The alloy we used is stainless steel 316L. The spatial and temporal simulation resolution are $\Delta x = \Delta y = 2\mu m$ and $\Delta t = 1\mu s$, respectively. More details of our physical SLS model will be reported in a separate paper on multi-physical SLS simulation.

In addition to sintering model for simulating porosity evolution, generation of powder bed (i.e. initial porous structure) is simulated using a "rain" model [26]. The mean and standard deviation of the diameter of deposited powders are 25μm and 0.5μm, respectively. It should be pointed out that yNet in this study will be trained to simulate porosity evolution and thus replace physics-based sintering model. Powder bed generation model is a separate model that provides initial structure for physical sintering simulation and yNet-based sintering simulation.

## 3. Architecture design of yNet

Fig. 1 illustrates the designed architecture of yNet, which is basically composed of the main encoder-decoder based on deep convolutional network and a branch of multilayer perceptron (MLP). The encoder plays a role as feature extractor, which distills the original high-dimensional field into numbers of information-rich and reduced-size feature maps. MLP acts to expand the condition parameters into numbers of neurons that generate a high-dimensional embedding vector. The encoder and MLP can greatly alleviate weight mismatch of original inputs, thereby facilitating effective fusion of input signals at their ends. Decoder serves to correctly reconstruct the merged signal back to a meaningful and desired field through the deep deconvolution process. Note that, in the deep deconvolution process we also concatenate feature maps extracted during the early autoencoding process; see gray arrows in Fig. 1c. This technique is widely adopted in semantic segmentation networks, such as U-Net [27] and FCN [28], with the aim of improving segmentation details. In the case of field evolvement, this technique is expected to endow yNet the capability of easily capturing tiny or even no field evolvement in partial region. For example, there is no porosity evolution/change at vertical locations far from heat source. It thus requires yNet to efficiently reconstruct feature maps to original porosity structure therein.

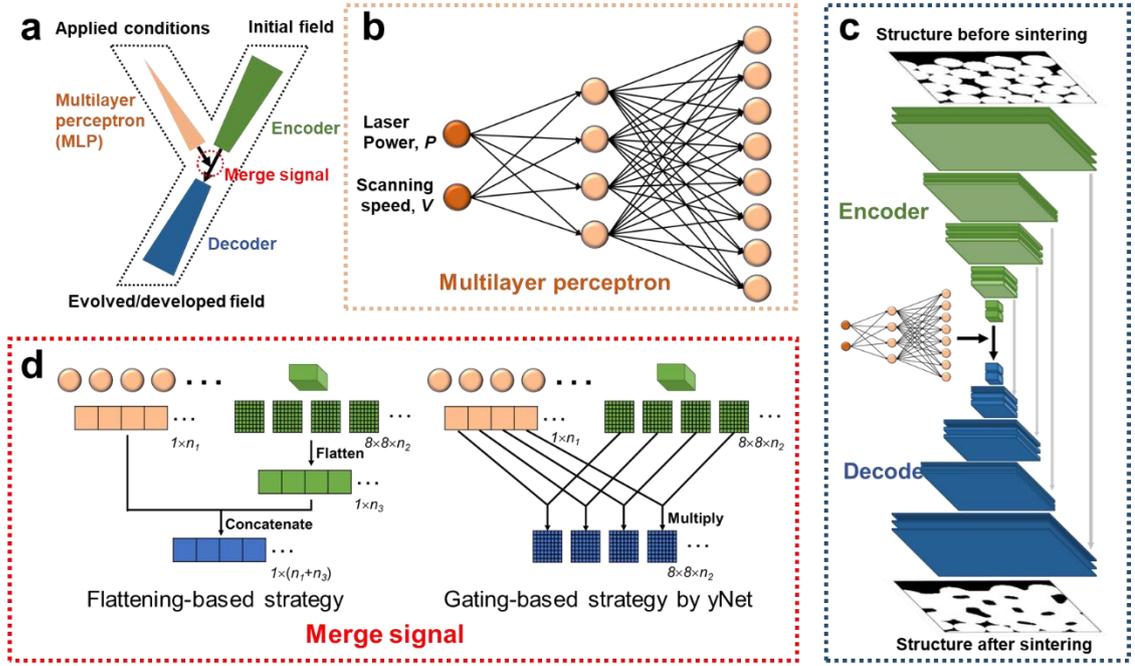

**Figure 1 The designed architecture of yNet for multi-input field-to-field regression.** (**a**) Overall architecture featuring a unique y-shape due to mixed inputs by incorporating condition-related parameters. (**b**) Multilayer perceptron (MLP) to expand condition-parameter inputs. (**c**) Encoder-decoder for automatic featurization of input field and reconstruction of merged signal to desired output field, respectively. (**d**) Comparison of flattening-based strategy with gating-based strategy used by yNet for merging signal at the end of MLP and encoder. Note that for the gating-based strategy, the final layer of MLP has to be designed with $n_1 = n_2$ due to one-to-one connection.

More specifically, for yNet instantiated in this study, the first layer of MLP is two condition inputs (i.e, laser power and scanning speed), the hidden layer is a fully connected layer with 128 neurons and the final layer 256 neurons. For the encoder, each green block represents a combo of Conv+Batch-Normalization+Relu operations and light green block means max-pooling. In this manner, encoder finally yields 256 8×8 feature maps, which are passed to a dropout layer with rate of 0.5 before merging with MLP signal. The decoder just has a somewhat mirrored topology of encoder, with each blue block representing a combo of Conv+Batch-Normalization+Relu operations and dark blue block up-sampling.

After building up the main architecture, the core problem is the proper manner in which two streams of input signal are merged. While multiple mixed inputs are not uncommon in various machine learning tasks, one merging strategy from the reservoir production predictor [29] (not image-to-image) is to flatten image, here represented by feature maps, into a high-dimensional vector, thus making it compatible and hence concatenable to the vectorial output of MLP; see Fig. 1d. Another merging strategy in image captioning model [30], which is not graphically illustrated here as it is very similar to the above one, is also flattening image first, however, strictly to a vector of the same size (i.e., $n_1 = n_3$), followed by adding two vectors (image and linguistic vectors) to successfully merge signals. Those flattening-based strategies might be effective for developing a simple label or continuous predictor, but will deteriorate the trainability and performance of yNet whose output is image/field. Because consistent signal flow of meaningful feature maps is desired throughout the encoding-decoding process, whereas the flattening operation in those strategies will interrupt the signal flow of feature maps. The merged signal in those manners becomes somewhat meaningless or less interpretable. A flattening-based strategy thus, if workable for image-to-image regression [31], might make yNet more like a brute-force regressor with poor interpretability and low trainability. What's more, flattening feature maps yields a large fully connected layer, which are notoriously parameter-intensive and can degrade the network performance (e.g. DeconvNet [32] versus SegNet [33]). Regarding the above facts, we propose to properly merge the signal using a one-to-one connection via multiplication inspired by the gating mechanism [34], thus without flattening feature maps. In this way, MLP actually turns as a signal modulator, which takes charge of signal strength of each feature map passed to decoder. The merged signal in this fashion is meaningful, still representing feature maps albeit with changed signal strength. Besides its full interpretability, the effectiveness of the gating-based merging strategy also lies in the fact of those neurons (or basically the applied conditions) interacting with respective feature map and thus the initial field behind in a direct, neat and therefore strong manner. We posit that, through training, MLP can learn to precisely manipulate the initial field represented by high-level feature maps into the developed field (preliminarily in the form of gated feature maps) for any given condition; the decoder then appropriately reconstruct the gated feature maps into the realistic developed field.

## 4. Results

### 4.1. Training and testing of yNet

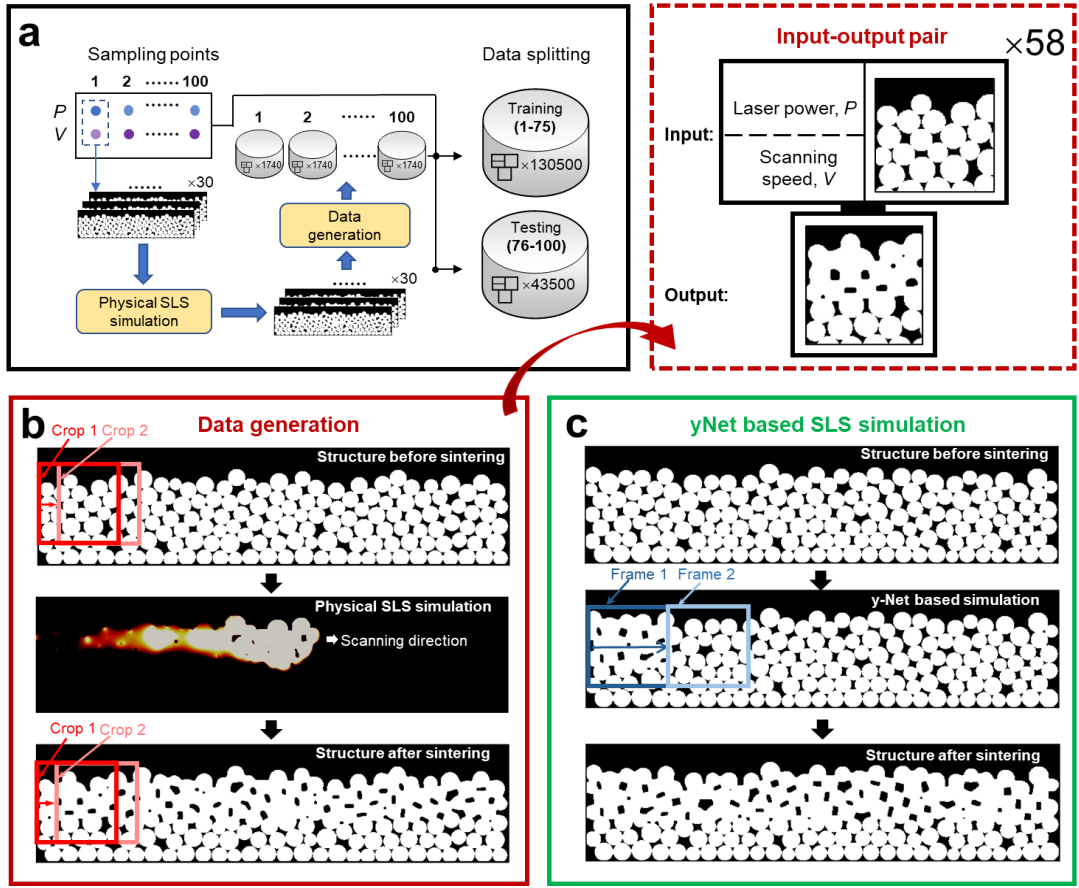

**Figure 2 Data generation and y-Net based simulation with respect to long tracks.** (**a**) Overall flowchart of training and testing of yNet. (**b**) To generate usable data by y-Net, each long track is cropped into a series of 128×128 standard patches, with a 118-pixel overlap between two consecutive patches as the cropped structure would change notably every 10 pixels. The same cropping strategy is applied to structures after sintering. As shown, a full input-output pair consists of applied laser conditions, a patch of structure before sintering and its corresponding structure after sintering. A 1400 mm or 700-pixel long track here would give us 58 pairs, which indicates a total of 1740 pairs obtained at each sampling point. (**c**) During yNet-based SLS simulation of a long track, we accordingly adopted a sliding inference window (but in a non-overlapped manner), until patch-wise completion of transforming the entire track into the sintered state. In this way, simulation of a 700-pixel long track would take 6 frames of inferences by yNet. Note that the top of the cropping and inference window are consistently in line with that of the powder bed. The 128×128 window is designed in this study to well accommodate the maximum potential sintering depth.

While neural network itself is physics-free, the training dataset does need to be sufficiently large to feed it enough physical knowledge. Regarding this, we generate 100 sampling points for laser power and scanning speed using Latin Hypercube sampling method [35], and then perform SLS simulation of 30 1400mm-long track of powder beds randomly generated at each point; see Fig. 2a. We repeatedly simulate 30 1400mm-long tracks simply because it is computationally much easier than simulating a single super-long (say 42000mm) track. A natural problem would then arise since we often deal with field of variable and large size in an engineering problem, as is the current case. However, the image/field processed by a neural network is usually limited to a small square of predefined size, as shown in Fig. 1. We thus adopt the data generation and simulation scheme as illustrated in Fig. 2b-c, to convert our engineering problem to a standard machine learning problem. Consequently, we get a dataset of 174000 input-output pairs in total. They are split into training and testing dataset based on laser power and scanning speed sample points, i.e., 75 conditions for training and 25 conditions for testing. Since all powder beds are also randomly generated, we guarantee zero replica between training and testing datasets in terms of either condition or structure. It is noted that while physical simulations and training of yNet are carried out on computing clusters, all SLS simulations

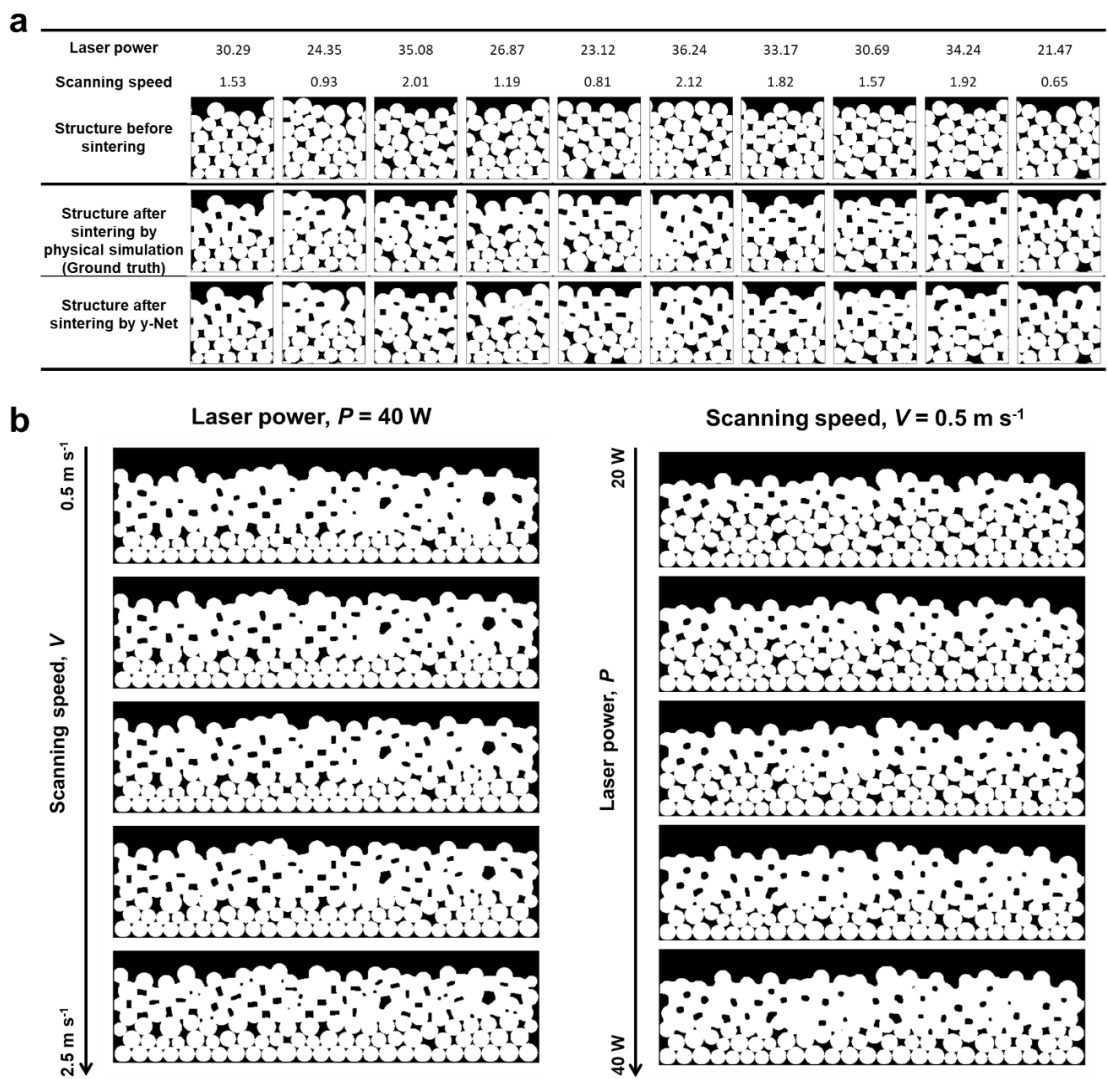

**Figure 3 Testing results of yNet. (a)** 10 testing results randomly picked from the 43500 testing results. Here yNet basically simulates the evolution of porous structure (described by phase field) under different laser conditions. **(b)** yNet-based systematic study of condition parameters to confirm that yNet has successfully learned the effect of conditions on field evolvement. It can be seen that, through end-to-end training, yNet has correctly learned the knowledge of increasing sintering depth and strength with decreasing scanning speed and increasing laser power. Note that since the sintering depth and strength may fluctuate with local structure, here we simulate long tracks to show the general sintering depth and strength under different conditions.

based on the trained yNet are performed on laptop to illustrate its computational friendliness.

We train yNet for 20 epochs with a mini-batch size of 2 and with cross entropy loss function, by using Adam Optimizer with learning rate = 0.001, $\beta_1$ = 0.9 and $\beta_2$ = 0.999. The weights with the smallest validation loss are saved and used in the remaining parts. Fig. 3a shows 10 randomly picked testing results. It can be seen that yNet results resemble closely to physics-based simulations and only small morphological error exists due to the extremely complicated evolving kinetics in the current case. As mentioned early, evolving kinetics of porosity evolution in SLS process is spatially non-uniform. Moreover, the evolving kinetics associated with sintering depth not only depends on laser power and scanning speed, but also fluctuates with local structure. In spite of the above fact, overall yNet can still accurately simulate field evolvement. Particularly noted is the surface that explicitly shows the performance of yNet on simulating evolution of complex geometry. Fig. 3a shows that yNet can perfectly simulate the evolution of complex surface structures. It should be pointed out that we currently take into account dominant condition parameters, i.e. laser power and scanning speed, but there are other condition parameters of secondary importance. Specifically, due to inherent limit of neural network, the adopted patch-wise simulation of yNet would give rise to additional condition parameters, because each patch is blind to neighboring structures. The current yNet can be thus further improved by quantifying such neighboring and environmental effects as condition parameters.

Since laser power and scanning speed in testing data are random, we also use the trained yNet to systematically study each condition parameter to clearly show the successful incorporation of their effects; see Fig 3b. This is the most distinctive capability of yNet over existing field-to-field regression models. In fact, the empirical knowledge of increasing sintering strength with higher laser power and lower scanning speed is easy for human to summarize by observing dataset. However, it is still quite interesting for yNet to have learned it through end-to-end training, thus 'self-aware' of sintering strength for any given condition, without resorting to any human-coded rule.

In addition to visual resemblance in Fig. 3a, by examining the global accuracy (i.e. percentage of correct pixels) for all 42750 testing results, yNet achieves an as high as 99.13% similarity to ground truth in average. All of the above qualitative and quantitative results prove the successful development of yNet for multi-input field-to-field regression tasks in engineering.

### 4.2. yNet boosted extreme-scale simulation

As a light-weight substitute of physics-based model, the trained yNet can help solve important engineering problems in need of large- or even extreme-scale simulation. For illustration purpose, here we present an as-of-yet

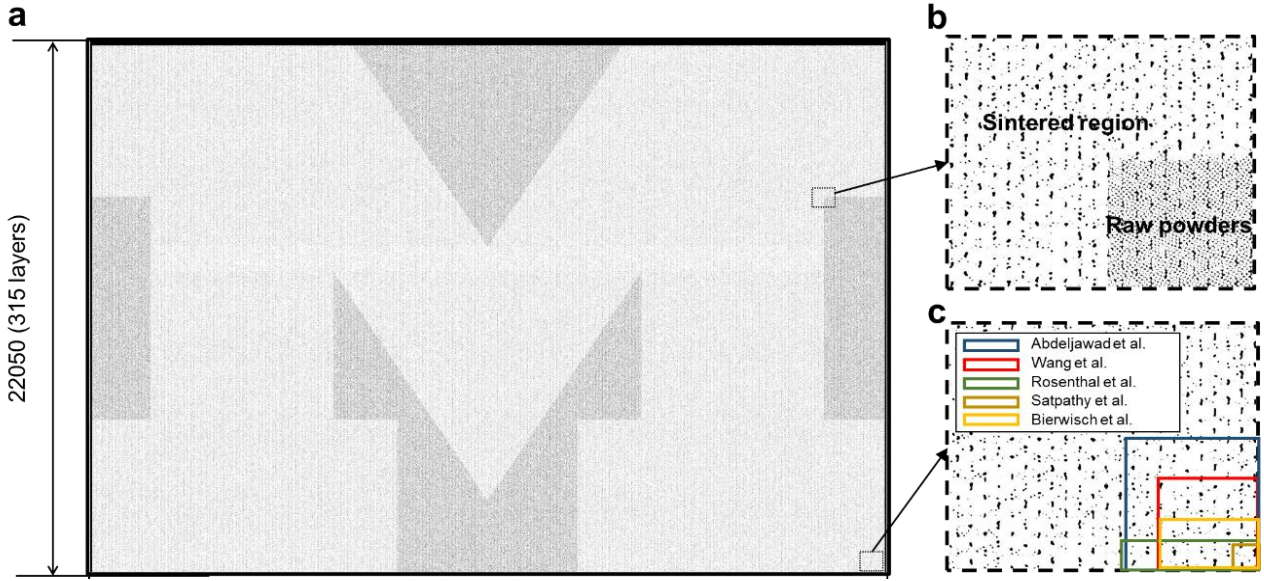

**Figure 4 yNet boosted porosity simulation of a full SLS component.** (**a**) y-Net based porosity simulation of a 315-layer high SLS component - block "M" logo of University of Michigan (35416×22050 pixels). (**b**) Magnified image showing the sintering details of a fraction of the SLS component. We point out that computational cost of powder bed generation increases exponentially with powder bed length. For memory and efficiency consideration, we simplify powder bed generation during full-component simulation, by depositing a continuous series of mini-powder-beds till completing deposition of the entire layer. This numerically creates large pores as regularly observed in between adjunct mini-powder-beds. Once available, a more efficient powder bed generation model can be integrated with our yNet for better component-level simulation. (**c**) Relative simulation size of existing physical simulations to yNet-based component-level simulation. They are compared in terms of dimensionless size in pixels.

computationally challenging task: full-component simulation of porosity development in SLS.

Fig. 4a shows the porosity simulation result of a realistic SLS component with dimensions of 70.8×44.1 mm$^2$ (or 35416×22050 pixels). Due to its extremely large size, here we can only present a fraction of the detail of the whole simulation result; see Fig. 4b. It is known that AM simulation is often limited to a single layer or even a representative volume element (RVE); a practical simulation of the entire layer-by-layer AM process is rarely achieved so far. Fig. 4a represents one of the first achievements of component-level AM simulation to date. Fig. 4c summarizes the simulation size of some of existing 2D sintering simulation researches [36-40]. Although prior researches did not report computation time and computational devices used, Fig. 4c gives a rough yet straightforward insight into the massively boosted simulation capability by yNet (several orders of magnitude larger). yNet enabled component-level simulation would open up enormous possibilities in SLS researches and practices. For example, SLS has been intentionally used for fabricating porous structures with high permeability and bio-compatibility [41, 42]. The trained yNet permits extensive virtual experiments in the pre-production phase, thus greatly facilitating materials design and process planning for fabricating a SLS component with any desired porosity distribution. It also allows for the ideal input of a full-component structure for computational evaluation of structure-property-performance relationship, from a combined local and global perspective.

### 4.3. Computation time

Here we quantitatively show the computational advantage of yNet by looking into the computation time. As mentioned early, yNet based simulations throughout this study are purposely performed on a laptop (Intel Core i7-7500U CPU, NVIDIA GeForce GTX 950M GPU, 16G RAM). In that case, the inference ability is 202 frames or unit patches per second. Consequently, simulating a single-layer track like those in Fig. 3b by yNet are just a matter of tens of milliseconds. In striking contrast, the original physical model that we used will take a few hours (depending on laser scanning speed simulated), to complete the same task in the same computational setting. Of special note is the porosity simulation for the full SLS component, which is previously a formidable and oftentimes impossible task even for high-end computing facilities but now accomplishable in a few minutes by using a laptop.

### 5. Discussion

A notable feature of yNet is its light weight, which might be one of the most important factors that makes it potentially a revolutionary simulation tool. As demonstrated, this merit enables the trained yNet to be readily deployed and ran with limited computational demand, thus democratizing the capability of ultra-fast and large-scale simulation. There are three main contributing factors to the light-weight yet high-performance of yNet. First, it benefits from its ConvNet based nature, as ConvNets are known for its expertise in efficiently dealing with image-type data [2]. Second, it should be admitted that a specific type of field in engineering usually has less morphological and textural complexity or variability than various real-world objects processed in a CV task. Third, the light-weight and trainability of yNet are largely attributed to smart design of its architecture, especially at the critical signal merging point. The successful development of yNet emphasizes the importance of designing network architecture in an interpretable manner. This helps guide the rational design of parameter-efficient and high-trainability neural network.

The multi-input field-to-field regression capability of yNet is basically supported by the tremendous non-linearity of deep neural network. However, before solid demonstration, we reserve the bold claim that yNet in its current form is suitable for all kinds of multi-input field-to-field regression, such as field evolvements with strong discontinuity and stochastic ones. But as clearly shown, it should be generally applicable to a lot of condition-dependent, continuous field evolvements, either in a spatially united or non-uniform evolving manner. Here the continuity of field evolvement can guarantee that even strong field evolvement showing significant morphological and textural changes can be well discretized into a series of mild ones, thus becoming readily tractable by a yNet that works in a recurrent manner. yNet can also handle temporally-varying evolving kinetics by simply incorporating time as a condition parameter, since the "condition" input imparts yNet immense potential and versatility. yNet would especially excel at simulating various structural and morphological evolutions [14] featuring an evolving field of high contrast or textural simplicity. yNet boosted simulation of them will be of huge academic and industrial value, since those evolving phenomena are of fundamental importance in numerous areas, such as additive manufacturing [43, 44], casting [45], welding [46], lithium-ion battery [47], to name a few. It should be nonetheless reminded that for more complex structures, one may need to scale-up the as-presented yNet to the adequate level. We also note that yNet can be easily extended to 3D with simple modifications like adoption of 3D convolutions and 3D pooling operations. It is known that computational cost of physical simulation usually increases sharply for 3D, in which case yNet may become even more invaluable.

Finally, we admit that in this study the patch-wise simulation strategy of yNet for dealing with large and variable simulation domain takes advantage of globally uniform evolving kinetics of porosity evolution in SLS. That is, while evolving kinetics are non-uniform inside each patch (locally), the overall evolving kinetics among patches are quite similar. The later is indicated by, under a given condition, the similar sintering strength and depth along the length of the entire track. In case of globally non-uniform field evolvements, we might incorporate the global location of patches as an additional condition parameter, to take into account the global effects on evolving kinetics. This will again capitalize on the versatility of yNet due to its capability of incorporating effects of "condition" on field evolvement. That being said, other solutions are however still desired for yNet to better tackle simulation domain of large and variable dimensions.

## 6. Conclusion

In summary, we have proposed a multi-input deep convolutional network, yNet, which offers the multi-input field-to-field regression capability urgently demanded in engineering. It offers an analytical-model-cheap way to simulate various field evolvements across many disciplines, by treating them as a physics-free, pure (field-to-field) regression problem. Using the simulation of porosity development in selective laser sintering as an example, we have demonstrated its strength in precisely simulating condition-dependent field evolvement with even spatially non-uniform evolving kinetics. yNet should be widely applicable to simulating different structural/morphological evolutions and other continuous field evolvements in engineering. It can help break the longstanding computational curse of those simulations and solve related engineering problems with no effort. It might become a revolutionary simulation tool by democratizing the capability of ultra-fast and extreme-scale simulation.